\documentclass[aps,pre,twocolumn]{revtex4-2}

\pdfoutput=1

\usepackage{mathtools}	
\usepackage{amssymb}	
\usepackage[italicdiff]{physics}	

\usepackage{xcolor} 		
\usepackage[colorlinks, citecolor={blue!70!black}, urlcolor={blue!70!black}, linkcolor={red!70!black},hyperindex,breaklinks]{hyperref}	
\usepackage{cleveref}	
\usepackage{enumerate}	

\usepackage{graphicx}	
\usepackage{float}		

\newcommand{\beq}{\begin{equation}}
\newcommand{\eeq}{\end{equation}}

\newcommand{\la}{\langle}
\newcommand{\ra}{\rangle}
\newcommand{\w}{\omega}
\renewcommand{\th}{\theta}
\renewcommand{\d}{\partial}

\begin{document}

\title{Quantum Fisher Information for Different States and Processes in Quantum Chaotic Systems}

\author{Fernando Iniguez}
\email{finiguez@ucsb.edu}
\author{Mark Srednicki}
\email{mark@physics.ucsb.edu}
\affiliation{Department of Physics, University of California, Santa Barbara, CA 93106}

\date{\today}

\begin{abstract}
The quantum Fisher information (QFI) associated with a particular process 
applied to a many-body quantum system has been suggested as a diagnostic 
for the nature of the system's quantum state,
e.g., a thermal density matrix vs. a pure state in a system that obeys the eigenstate
thermalization hypothesis (ETH). We compute the QFI for both an energy eigenstate and 
a thermal density matrix for a variety of processes in a system obeying ETH, including
a change in the hamiltonian that is either sudden (a quench),
slow (adiabatic), or followed by contact with a heat bath. 
We compare our results with earlier results for a local unitary transformation.
\end{abstract}

\maketitle

\section{Introduction}\label{intro}

Quantum Fisher information (QFI; for an introduction and review, see \cite{QFIreview})
has recently received renewed attention as a diagnostic tool for understanding 
properties of quantum many-body systems \cite{QFIZoller,QFISilva}.
The QFI for a one-parameter family of density operators $\rho_\th$ is designated $F_\th$, and one of its key properties is that it sets the minimum uncertainty $\Delta\th$ in the value of $\th$ after an ideal measurement. 
Specifically, for a single ideal measurement, the Cram\'er--Rao bound 
is
\beq
(\Delta\th)^2 \ge \frac{1}{F_\th}.
\label{CR}
\eeq
This can be viewed as a generalized form of the uncertainty principle. 

In practice, we are most interested in density operators $\rho_\th$ whose $\th$ dependence is due to some 
specific experimental manipulation on a base $\rho_0$,
such as a local unitary transformation of the form $\rho_\th=U^\dagger_\th \rho_0 U_\th$ with
$U_\th = \exp(-i\th A)$ for some local hermitian operator $A$. 
This corresponds to an experimental set-up in which 
the experimenter has direct control of the physical quantity represented by $A$, such as a single qubit.

In \cite{QFIZoller}, the QFI for a system whose quantum state is a thermal density matrix that is subjected
to this type of local unitary transformation was expressed in terms of a particular dynamical susceptibility.
In \cite{QFISilva}, it was pointed out that, in a system that obeys ETH, the QFI for an energy eigenstate
differs from the the QFI for a thermal density matrix when both are subjected to the same 
local unitary transformation. This makes the QFI a useful theoretical tool for distinguishing 
pure and mixed states that are not distinguished by measurements of local observables. 
In this regard the QFI is comparable to the von Neumann entropy
and Renyi entropies of $\rho_\th$. 

In this work, in addition to the local unitary transformation described above, we consider three other 
possible experimental protocols performed on a quantum chaotic system: 
an adiabatic transformation,  in which the system's hamiltonian is very slowly changed; 
a quench, in which the system's hamiltonian is suddenly changed and the system is then allowed to evolve in time under the new hamiltonian (this case has been discussed previously in \cite{QuenchQFIex}); 
and a rethermalization, in which a system that was initially thermalized by contact with a heat bath has its hamiltonian changed, and then is put back in contact with the same heat bath. We compute the QFI for these transformations for an energy eigenstate and for a thermal density matrix 
(only the latter in the case of rethermalization).

\section{Quantum Fisher Information}\label{QFI}

The QFI is given formally by
\beq
F_\th = \Tr L^2_\th\rho_\th ,
\label{Fth}
\eeq
where $L_\th$ (the symmetric logarithmic derivative) is an operator that is defined implicitly via
\beq
\rho'_\th = {\textstyle\frac12}\bigl(L_\th\rho_\th+\rho_\th L_\th\bigr),
\label{Lth}
\eeq
where the prime denotes a derivative with respect to $\th$.
Eq.~(\ref{CR}) holds if the rank of $\rho_\th$ does not change as $\th$ is varied over a small range 
around its base value \cite{QFIRank}, which will always be the case in this work. 

A more explicit formula for the QFI follows from the spectral decomposition of $\rho_\th$,
\beq
\rho_\th = \sum_i p_i |i\ra\la i|,
\label{rhoth}
\eeq
where the states $\{|i\ra\}$ form an orthonormal and complete basis, and the probabilities
$p_i$ obey $0\le p_i\le1$ and $\sum_ip_i=1$. (These states and probabilities depend on $\th$, 
but we do not denote this explicitly.) We then have
\beq
F_\th =2\sideset{}{'}\sum_{ij}\frac{|\la i|\rho'_\th|j\ra|^2}{p_i+p_j},
\label{Fth2}
\eeq
where the prime on the sum means that terms for which $p_i+p_j=0$ (if any) are omitted.

Because of the linearity of quantum mechanics, in all cases of practical interest 
$\rho'_\th$ is linearly related to $\rho_\th$. For the four specific types of transformations
that we consider, this linear relation takes the form
\beq
\rho'_\th = i[B,\rho_\th] + D\rho_\th,
\label{rhopB}
\eeq
where $B$ and $D$ are hermitian operators with $[D,\rho_\th]=0$ and $\Tr D\rho_\th=0$.
We note that the normalization condition $\Tr\rho_\th=1$ implies $\Tr\rho'_\th=0$,
which is satisfied by Eq.~(\ref{rhopB}).
Using Eq.~(\ref{rhopB}) in Eq.~(\ref{Fth2}), we get
\beq
F_\th =2\sideset{}{'}\sum_{ij}\frac{(p_i-p_j)^2}{p_i+p_j}|\la i|B|j\ra|^2
+\sum_i p_i |\la i|D|i\ra|^2.
\label{F0}
\eeq

For a transformation that is unitary (which is the case for all but one of the transformations
we consider), we have $D=0$. In this case, an important relation obeyed by $F_\th$ is 
\beq
F_\th \le 4(\Delta B)^2,
\label{Fvar}
\eeq
where
\beq
(\Delta B)^2 = \Tr\rho_\th B^2 - (\Tr\rho_\th B)^2
\label{DB2}
\eeq
is the quantum variance in the expectation value of $B$ in the state $\rho_\th$.
Eq.~(\ref{Fvar}) becomes an equality if and only if $\rho_\th$ is a pure state.
Thus the QFI of a mixed state is strictly less than that of a pure state with the same
quantum variance of $B$. 

Note that Eqs.~(\ref{CR},\ref{Fvar},\ref{DB2}) yield $(\Delta\th)(\Delta B)\ge 1/2$, 
which is a more recognizable form of the uncertainty principle.

\section{Experimental protocols}\label{prot}

We consider four types of experimental protocols that transform an initial reference state
$\rho_0$ to $\rho_\th$.

{\it Local unitary.} We set
\beq
\rho_\th = e^{i\th A}\rho_0 e^{-i\th A},
\label{lu}
\eeq
where $A$ is a dimensionless local hermitian operator. 
This is the form of $\rho_\th$ that is treated in \cite{QFISilva}.

{\it Adiabatic.}
The hamiltonian is slowly changed from $H$ to 
\beq
H_\th =H+\th\mu A,
\label{Hth}
\eeq
where $\mu$ is a constant with dimensions of energy.  (This constant could be absorbed into either
$\th$ or $A$, but we prefer to keep both these quantities dimensionless to facilitate comparison
of different transformation protocols.)
As $H$ is slowly changed from $H$ to $H_\th$, an eigenstate $|\alpha\ra$ of $H$ evolves adiabatically 
to an eigenstate $|\alpha\ra_\th$ of $H_\th$.
Hence, an initial density operator $\rho_0$ evolves to
\beq
\rho_\th = \sum_{\alpha,\beta}|\alpha\ra_\th \la\alpha|\rho_0|\beta\ra\la\beta|_\th .
\label{ad}
\eeq
This is a unitary transformation.

{\it Quench.}  
The hamiltonian is suddenly changed from $H$ to $H_\th$, Eq.~(\ref{Hth}),
and the system then evolves unitarily under $H_\th$ for a time $t$. This yields
\beq
\rho_\th = e^{-iH_\th t}\rho_< e^{iH_\th t},
\label{quinit}
\eeq
where $\rho_<$ is the state of the system just before the quench at $t=0$. This differs
from $\rho_0$, which is defined as the state of the system with $\th=0$ at time $t$.
In terms of $\rho_0$, Eq.~(\ref{quinit}) becomes
\beq
\rho_\th =e^{-iH_\th t}e^{iHt}\rho_0 e^{-iHt} e^{iH_\th t}. 
\label{qu}
\eeq
This is a unitary transformation.

{\it Rethermalization.}
For this protocol, we specialize to the case that the initial state is a thermal
density operator for the hamiltonian $H$ at an inverse temperature $\beta$,
\beq
\rho_0 = Z_0^{-1}e^{-\beta H},
\label{rho0}
\eeq
where $Z_0=\Tr e^{-\beta H}$.
The density operator $\rho_\th$ is then taken
to be a thermal density operator at the same inverse temperature $\beta$,
but now with hamiltonian $H_\th$, Eq.~(\ref{Hth}),
\beq
\rho_\th = Z_\th^{-1}e^{-\beta H_\th},
\label{rhob}
\eeq
where $Z_\th=\Tr e^{-\beta H_\th}$.
This protocol corresponds to putting the system (with hamiltonian $H$) in contact
with a heat bath at inverse temperature $\beta$, then changing the hamiltonian to
$H_\th$, and then putting the system back into contact with the same heat bath.
This is not a unitary transformation.

We now compute $\rho'_0$ for each of these transformations (for an arbitrary $\rho_0$)
and identify the operators $B$ and $D$ in Eq.~(\ref{rhopB}), 
which then yields the QFI via Eq.~(\ref{F0}). 

For any unitary transformation (which includes our local unitary, adiabatic, and quench
transformations), we have
\beq
D=0 
\quad\mbox{(local unitary, adiabatic, quench)}.
\label{Ceq0}
\eeq

For the local unitary transformation of Eq.~(\ref{lu}), taking the derivative with respect to $\th$ and comparing to Eq.~(\ref{rhopB}) yields
\beq
B=A 
\quad\mbox{(local unitary)}.
\label{BeqA}
\eeq

For the adiabatic transformation of Eq.~(\ref{Hth}), an eigenstate $|\alpha\ra_\th$ of $H_\th$ is found from Rayleigh-Schrodinger perturbation theory to be
\beq
|\alpha\ra_\th = |\alpha\ra + \th\mu\sideset{}{'}\sum_\gamma \frac{A_{\gamma\alpha}}{E_\alpha-E_\gamma}|\gamma\ra + O(\th^2),
\label{alphath}
\eeq
where the prime means that $\gamma=\alpha$ is omitted, and $|\alpha\ra$ is an eigenstate of $H$
with eigenvalue $E_\alpha$,
\beq
H|\alpha\ra = E_\alpha|\alpha\ra.
\label{HaEa}
\eeq
Using Eq.~(\ref{alphath}) in Eq.~(\ref{ad}) and taking the derivative with respect to $\th$, we get Eq.~(\ref{rhopB}) with the matrix elements of $B$ (in the energy eigenstate basis) given by
\beq
B_{\alpha\beta} = 
\begin{cases} 
0 &\mbox{if } \alpha=\beta \\ 
{\displaystyle\frac{i\mu A_{\alpha\beta}}{E_\alpha-E_\beta}} & \mbox{if } \alpha\ne\beta \end{cases} 
\quad\mbox{(adiabatic)}.
\label{Babad}
\eeq
We note that these matrix elements of $B$ are the same as the matrix elements of the
adiabatic gauge potential, a quantity introduced in \cite{AGP} 
as a diagnostic tool for quantum chaos.

The quench transformation of Eq.~(\ref{qu}) is unitary, and we can express $B$ via
\beq
B =-i\frac{\d}{\d\th}e^{-iH_\th t}e^{iHt}\Big|_{\th=0}
\label{Bqu} 
\eeq
As shown in Appendix \ref{quenchderivation}, this yields the matrix elements
\beq
B_{\alpha\beta} = 
\begin{cases} 
-\mu t A_{\alpha\alpha} &\mbox{if } \alpha=\beta \\ 
\noalign{\smallskip}
{\displaystyle\frac{1-e^{-i(E_\alpha-E_\beta)t}}{E_\alpha-E_\beta}i\mu A_{\alpha\beta}} & \mbox{if } \alpha\ne\beta \end{cases} 
\quad\mbox{(quench)}.
\label{Babqu}
\eeq

For the rethermalization transformation of Eq.~(\ref{rhob}), 
we find, as shown in Appendix \ref{rethermalderivation},
\beq
B_{\alpha\beta} = 
\begin{cases} 
0 &\mbox{if } \alpha=\beta \\ 
\noalign{\smallskip}
{\displaystyle\frac{i \mu A_{\alpha\beta}}{E_\alpha-E_\beta}} & \mbox{if } \alpha\ne\beta \end{cases} 
\quad\mbox{(rethermalization)}
\label{Babret}
\eeq
and
\beq
D_{\alpha\alpha} = -\beta\mu\bigl(A_{\alpha\alpha}-\la A\ra\bigr)
\quad\mbox{(rethermalization)},
\label{Daa} 
\eeq
where 
\beq
\la A\ra = \Tr\rho_0 A .
\label{laAra}
\eeq

In addition to specifying the experimental protocols, we must also specify the
initial state. As in \cite{QFISilva}, we compare and contrast the results for an initial 
energy eigenstate and a thermal density operator with the same energy. 
Our results can be straightforwardly generalized to other classes of initial states,
both pure and mixed; we comment briefly on this in the conclusions.

We also note that for any $\rho_0$ that is diagonal in the energy basis, 
the diagonal elements $B_{\alpha\alpha}$ of $B$ drop out of 
the right-hand side of Eq.~(\ref{rhopB}), and hence do not affect the value of the QFI.

\section{Review of ETH}\label{revETH}

We assume that the system of interest is a closed, finite, chaotic many-body 
system (with $N\gg 1$ degrees of freedom). Our working definition of {\it chaotic} is that 
each few-body observable $A$ obeys the eigenstate thermalization hypothesis (ETH),
which states that the matrix elements of $A$ in the energy basis take the form \cite{Mark98}
\beq
A_{\alpha\beta} = \mathcal{A}(E)\delta_{\alpha\beta} + e^{-S(E)/2}f(E,\w)R_{\alpha\beta},
\label{Aab}
\eeq
where $E=(E_\alpha+E_\beta)/2$ is the average energy of the two eigenstates, $\w=E_\alpha-E_\beta$ is the energy difference,
$S(E)$ is the thermodynamic entropy (logarithm of the density of states) at energy $E$,
${\cal A}(E)$ and $f(E,\w)$ are smooth, real functions of their arguments, with $f(E,\w)=f(E,-\w)$, 
and $R_{\alpha\beta}$ is a hermitian matrix of erratically varying elements, 
with overall zero mean and unit variance in local ranges of $E$ and $\w$.
The function $f(E,\w)$ can be related to the dynamical susceptibility of $A$ \cite{QFISilva}.
We take $E$ to be an extensive quantity ($E\sim N$) 
and $\w$ to be an intensive quantity ($\w\sim 1$).
In accord with this, 
we assume that the initial state $\rho_0$ yields an expectation value of the energy
that is extensive, 
\beq
E=\Tr\rho_0 H \sim N,
\label{ETrH}
\eeq
and a quantum energy uncertainty that is sub-extensive,
\beq
\Delta E = [\Tr\rho_0(H-E)^2]^{1/2} \sim N^\nu, \quad \nu<1.
\label{DE}
\eeq
We note that for a thermal state, $\nu=1/2$.

\section{Computing QFI}\label{CQFI}

We begin by considering an initial energy eigenstate, $\rho_0=|\alpha\ra\la\alpha|$, and a local unitary transformation, 
Eq.~(\ref{lu}), which yields Eq.~(\ref{BeqA}). Since the initial state is pure, Eq.~(\ref{Fvar}) holds as an equality. Using Eq.~(\ref{DB2}) and inserting a complete set of energy eigenstates, we have
\beq
F_0 = 4\sum_{\beta\ne\alpha}|A_{\alpha\beta}|^2. \quad\mbox{(local unitary)}.
\label{F01} 
\eeq
We now use the ETH ansatz of Eq.~(\ref{Aab}).
We replace $|R_{\alpha\beta}|^2$ by its statistical average of 1
over a small range of $E_\beta$, and convert the sum over $\beta$ 
to an integral over $E_\beta$; this integral includes 
a density-of-states factor of $\exp S(E_\beta)$.
We then change the integration variable to $\w=E_\alpha-E_\beta$.
The result is
\beq
F_0 = 4\int_{-\infty}^{+\infty}\!\! d\w\,
e^{S(E_\alpha-\w)-S(E_\alpha -\w/2)} 
|f(E_\alpha{+}\w/2,\w)|^2.
\label{F02}
\eeq
Treating $E_\alpha$ as extensive and $\w$ as intensive, we can Taylor expand
the $S(E)$ factors using $\beta \coloneqq S'(E_\alpha)$, where $\beta$
is the inverse temperature of the system when the energy is $E_\alpha$.
We can also neglect the shift of $E_\alpha$ in the first argument of $f(E,\w)$.
Finally, we can use the fact that $f(E,\w)$ is an even function of $\w$.
The result is
\beq
F_0 = 4\int_{-\infty}^{+\infty}\!\! d\w
\cosh\Bigl(\frac{\beta\w}{2}\Bigr) |f(E,\w)|^2. \quad\mbox{(local unitary)},
\label{F03}
\eeq
is in agreement with \cite{QFISilva}.
We have dropped the $\alpha$ index on $E$ for notational simplicity.
At small $\w$, we generally expect $f(E,\w)$ to approach a nonzero constant,
and at large $\w$, $|f(E,\w)|^2$ goes to zero 
faster than $\exp(-\beta|\w|/2)$ \cite{ChaiMark2019}.
Hence this integral converges. 

For the other two transformation protocols that we consider 
in the case of an initial energy eigenstate (adiabatic and quench), 
we express our results in terms of a function $K(\w)$, defined via 
\beq
F_0 = 4\int_{-\infty}^{+\infty}\!\! d\w\,K(\w)\,
\cosh\Bigl(\frac{\beta\w}{2}\Bigr) |f(E,\w)|^2.
\label{F0K}
\eeq
As we have already seen, for an energy eigenstate (es) and a local unitary transformation (lu), we have
\beq
K_{\mathrm{es,lu}}(\w)=1.
\label{Klu}
\eeq

For an initial energy eigenstate and adiabatic transformation (ad), we can deduce $K(\w)$ by comparing Eq.~(\ref{Babad}) with Eq.~(\ref{BeqA}). We see that
\beq
K_{\mathrm{es,ad}}(\w)=\frac{\mu^2}{\w^2}.
\label{Kad}
\eeq
Since we generically expect $f(E,\w)$ to approach a nonzero constant as $\w\to 0$, 
the integral in Eq.~(\ref{F0K}) diverges at low $\w$. 
There is a lower cutoff at the mean level spacing
$\Delta \sim \exp[-S(E)]$, so in this case the QFI is exponentially large,
$F_0 \sim \exp S(E)$. Note, however, that for the transformation to be truly
adiabatic, with negligible possibility of changing energy levels, it must be done over an equally exponentially large time. 

For a quench (qu), we get $K(\w)$ by comparing Eq.~(\ref{Babqu}) with Eq.~(\ref{BeqA}). We find
\beq
K_{\mathrm{es,qu}}(\w)=4\frac{\mu^2}{\w^2}\sin^2\Bigl(\frac{\w t}{2}\Bigr).
\label{Kqu}
\eeq
In this case, $K_{\mathrm{es,qu}}(\w)\to \mu^2 t^2$ as $\w\to0$, and so the integral in Eq.~(\ref{F0K}) does not diverge at low $\w$.
In the limit of large $t$, following the standard procedure used to derive 
Fermi's Golden Rule for a transition rate, we can make the replacement
\beq
\frac{1}{\w^2}\sin^2\Bigl(\frac{\w t}{2}\Bigr)\to \frac{\pi}{2}|t|\delta(\w).
\label{larget}
\eeq
This implies that at late times after the quench, the QFI grows at a constant rate of
\beq
\frac{dF_0}{dt} = 8\pi\mu^2 |f(E,0)|^2.
\label{dFdt}
\eeq
After an exponentially long time, due to the discreteness of the energy levels,
this will saturate at the exponentially large value
$F_0 \sim \exp S(E)$ that we found for an equally long adiabatic transformation.

We now consider a thermal initial state, Eq.~(\ref{rho0}). 
This is diagonal in the energy basis, with
$p_\alpha = e^{-\beta E_\alpha}/Z_0$. Hence Eq.~(\ref{F0}) becomes
\beq
F_0 =2\sideset{}{'}\sum_{\alpha,\beta}\frac{(p_\alpha-p_\beta)^2}{p_\alpha+p_\beta}|B_{\alpha\beta}|^2.
\label{F06}
\eeq

For a local unitary transformation we have $B=A$. We use the ETH ansatz of Eq.~(\ref{Aab}),
replace $|R_{\alpha\beta}|^2$ by its average of 1
over small energy ranges, and convert the sums to 
integrals over $E_\alpha$ and $E_\beta$, including
a density-of-states factor of $\exp S$ for each.
We then change the integration variables to $E=(E_\alpha+E_\beta)/2$, $\w=E_\alpha-E_\beta$.
The result is
\beq
F_0 = \frac{4}{Z_0}\int_{E,\w} e^{S(E)-\beta E} \,\frac{\sinh^2(\beta\w/2)}{\cosh(\beta\w/2)}|f(E,\w)|^2,
\label{F07}
\eeq
where $\int_{E,\w}\coloneqq \int_0^\infty dE\int_{-\infty}^{+\infty}d\w$.
Performing the integral over $E$ by Laplace's method fixes
the value of $E$ at the solution of $S'(E)=\beta$,
and yields a factor of $Z_0$. Hence for a thermal initial state and a local unitary transformation, we get Eq.~(\ref{F0K}) with
\beq
K_{\mathrm{th,lu}}(\w) = \tanh^2\Bigl(\frac{\beta\w}{2}\Bigr),
\label{Kthlu}
\eeq
The integral in Eq.~(\ref{F0K}) then converges at both high and low $\w$. 
This result is in agreement with \cite{QFISilva}.

The relative factor between $K_{\mathrm{es,lu}}$ and $K_{\mathrm{th,lu}}$
comes solely from the different spectrum of $p_\alpha$, and hence the same
relation holds for the other transformations,
\beq
K_{\mathrm{th},i}(\w) = \tanh^2\Bigl(\frac{\beta\w}{2}\Bigr)K_{\mathrm{es},i}(\w),
\label{Kthi}
\eeq
where $i=\mathrm{lu}, \mathrm{ad}, \mathrm{qu}$.
Hence for a thermal initial state and an adiabatic transformation, we have
\beq
K_{\mathrm{th,ad}}(\w) =\frac{\mu^2}{\w^2}\tanh^2\Bigl(\frac{\beta\w}{2}\Bigr).
\label{Kthad}
\eeq
For a  a thermal initial state and a quench, we have
\beq
K_{\mathrm{th,qu}}(\w) = 4\frac{\mu^2}{\w^2}\tanh^2\Bigl(\frac{\beta\w}{2}\Bigr)\sin^2\Bigl(\frac{\w t}{2}\Bigr). 
\label{Kthqu}
\eeq
For the adiabatic and quench transformations, 
the integral in Eq.~(\ref{F0K}) converges at both high and low $\w$.
For the quench in the limit of large $t$, we can make the replacement
\beq
\sin^2\Bigl(\frac{\w t}{2}\Bigr)\to \frac{1}{2}
\label{larget2}
\eeq
instead of Eq.~(\ref{larget}); the difference is due to the differing
behavior of the integrand at low $\w$.
The QFI in the case of a quench then
approaches a constant value, equal to twice its value in the adiabatic case.

For rethermalization, a comparison of Eq.~(\ref{Babret}) with Eq.~(\ref{BeqA}) yields
\beq
K_{\mathrm{th,re}}(\w) =\frac{\mu^2}{\w^2}\tanh^2\Bigl(\frac{\beta\w}{2}\Bigr),
\label{Kthreth}
\eeq
which is the same as the adiabatic case for a thermal initial state.

\section{Conclusions and outlook}
We have computed the QFI for several different possible experimental protocols performed
on a quantum many-body system that obeys the eigenstate thermalization hypothesis (ETH).
The results are expressed in terms of a form-factor $K(\w)$ via Eq.~(\ref{F0K}).
The difference between an energy eigenstate and a thermal density matrix is the same
for all protocols, and is given by Eq.~(\ref{Kthi}), in agreement with the results of \cite{QFISilva}.

The most dramatic difference occurs for an adiabatic transformation, where the QFI is
exponentially larger for a thermal density matrix than for an energy eigenstate (but the
transformation must be made exponentially slowly). After a quench, the QFI grows linearly
with time for a thermal density matrix while it remains constant for an energy eigenstate.

The results for an energy eigenstate will also hold for ``typical'' pure state with a non-extensive
energy uncertainty. By ``typical'', we mean a state whose expansion coefficients in the energy-eigenstate
basis are uncorrelated with the matrix elements of the transformation operator $A$. 

For a similarly ``typical'' mixed state, the results will depend on the details of the spectrum of 
probabilities $p_i$ in the diagonalizing basis.

We have thus seen that the QFI is a valuable theoretical diagnostic tool. Its value distinguishes between
a self-thermalized pure state in a quantum-chaotic system and a true thermal density matrix,  
and furthermore distinguishes among different experimental measurement protocols, allowing for
a precise characterization of the measurement uncertainty in these different situations.

\begin{acknowledgments}
The work of F.I. was supported by an NSF Graduate Research Fellowship under Grant No.~2139319 and funds from the University of California.
\end{acknowledgments}

\clearpage
\onecolumngrid
\appendix
\section{Matrix elements of B for a quench}
\label{quenchderivation}
We start with the general identity \cite{Wilcox}
\beq
\frac{\d}{\d\th}e^{-iH_\th t}
= -i\int_0^t dt' e^{-iH_\th t'}\biggl(\frac{\d H_\th}{\d\th}\biggr)e^{-iH_\th(t-t');}.
\label{BCH}
\eeq
Setting $H_\th = H + \th\mu A$ and using Eq.~(\ref{BCH}) in Eq.~(\ref{Bqu}), we get
\beq
B =
-\int_0^t dt' e^{-iHt'}\mu Ae^{iHt'}
\label{B2}
\eeq
for the case of a quench. Sandwiching Eq.~(\ref{B2}) between two different 
energy eigenstates $\la\alpha|$ and $|\beta\ra$, 
and performing the integral over $t'$, we obtain the off-diagonal matrix elements
\begin{equation}
    B_{\alpha \beta} = \frac{1-e^{-i (E_\alpha-E_\beta)} t}{E_\alpha-E_\beta}i\mu A_{\alpha\beta}.
\label{Babqu2}
\end{equation}
Taking the limit $E_\beta\to E_\alpha$ yields the diagonal matrix elements
\begin{equation}
    B_{\alpha \alpha} = -\mu t A_{\alpha \alpha}.
    \label{Baaqu}
\end{equation}
This verifies Eq.~(\ref{Babqu}) in the main text.

\section{Matrix elements of B and D for rethermalization}
\label{rethermalderivation}
We begin by taking the derivative with respect to $\theta$ of $\rho_\theta$ as given by Eq.~(\ref{rhob}),
\begin{equation}
\frac{\d}{\d\theta}\rho_\theta = \frac{1}{Z_\theta} \frac{\d}{\d\theta}e^{-\beta H_\theta} -\frac{1}{Z^2_\theta}\frac{\d Z_\theta}{\d \theta} e^{-\beta H_\theta}.
\end{equation}
Applying Eq.~(\ref{BCH}) with $t\to -i\beta$ and $t'\to -i\beta'$ and setting $\theta=0$ yields
\begin{equation}
    \rho'_0
    = -\frac{1}{Z_0} \int_0^\beta d\beta' e^{-\beta'H}\mu Ae^{-(\beta -\beta')H} 
    + \frac{1}{Z_0^2}e^{-\beta H} \Tr\int_0^\beta d\beta' e^{-\beta'H}\mu Ae^{-(\beta -\beta')H}.
    \label{b2}
\end{equation}
Using the cyclic property of the trace in the second term, we get
\begin{equation}
    \rho'_0 
    = -\frac{1}{Z_0} \int_0^\beta d\beta' e^{-\beta'H}\mu Ae^{-(\beta -\beta')H} 
     + \beta\mu\la A\ra\rho_0 ,
\label{drdt0}
\end{equation}
where $\la A\ra = \Tr \rho_0 A$.

Sandwiching Eq.~(\ref{drdt0}) between two different energy eigenstates $\la\alpha|$ and $|\beta\ra$ yields zero from the second term since $\rho_0$ is diagonal in the energy basis.
Performing the integral over $\beta'$ in the first term then results in the off-diagonal matrix elements
\begin{equation}
   \la\alpha|\rho'_0|\beta\ra 
   = \frac{1}{Z_0}\bigl(e^{-\beta E_\alpha}-e^{-\beta E_\beta}\bigr)
    \frac{\mu A_{\alpha \beta}}{E_\alpha - E_\beta}.
   \label{b4}
\end{equation}
Using $\rho_0 = Z_0^{-1}e^{-\beta H}$, we have
\begin{equation}
   \la \alpha |[B,\rho_0]|\beta \ra 
   =-\frac{1}{Z_0} \bigl(e^{-\beta E_\alpha }- e^{-\beta E_\beta}\bigr)B_{\alpha \beta}.
   \label{b6}
\end{equation}
Using Eq.~(\ref{rhopB}) and comparing with Eq.~(\ref{b4}), we identify the off-diagonal matrix elements of $B$ as those of Eq.~(\ref{Babret}). 

The diagonal elements of $B$ can be set to zero,
since they do not appear in Eq.~(\ref{F0}) for any $\rho_0$ that is diagonal in the energy basis, which is the only case we consider for rethermalization. 

Sandwiching Eq.~(\ref{drdt0}) between identical energy eigenstates $\la\alpha|$ and $|\alpha\ra$ yields the limit of Eq.~(\ref{b4}) as $E_\beta\to E_\alpha$, plus the expectation value of the
second term on the right-hand side of Eq.~(\ref{drdt0}),
\beq
\la\alpha|\rho'_0|\alpha\ra 
   = -\frac{1}{Z_0}e^{-\beta E_\alpha}\beta\mu\bigl(A_{\alpha\alpha}-\la A\ra\bigr).
   \label{b7}
\eeq
We also have, using $[D,\rho_0]=0$,
\beq
\la\alpha|D\rho_0|\alpha\ra = \frac{1}{Z_0}e^{-\beta E_\alpha}D_{\alpha\alpha}.
\label{Daa2}
\eeq
Comparing Eqs.~(\ref{b7}) and (\ref{Daa2}) yields Eq.~(\ref{Daa}).

\clearpage

\bibliography{QFI-refs.bib}

\end{document}